\documentclass[letterpaper,english,reprint, aps,prl.floatfix,superscriptaddress]{revtex4-1}
\usepackage[T1]{fontenc}
\usepackage[latin9]{inputenc}
\usepackage{amsmath}
\usepackage{amssymb}
\usepackage{graphicx}

\makeatletter

\pdfpageheight\paperheight
\pdfpagewidth\paperwidth

\makeatother

\usepackage{babel}
\begin{document}
\title{A Multiorbital Quantum Impurity Solver for General Interactions and
Hybridizations}
\author{Eitan Eidelstein}
\affiliation{Department of Physics, NRCN, P.O. Box 9001, Beer Sheva 84190, Israel}
\affiliation{School of Chemistry, Tel Aviv University, Tel Aviv 69978, Israel}
\author{Emanuel Gull}
\affiliation{Department of Physics, University of Michigan, Ann Arbor, Michigan
48109, USA}
\affiliation{Center for Computational Quantum Physics, Flatiron Institute, New
York, New York 10010, USA}
\author{Guy Cohen}
\affiliation{School of Chemistry, Tel Aviv University, Tel Aviv 69978, Israel}
\date{\today}
\begin{abstract}
We present a numerically exact Inchworm Monte Carlo method for equilibrium
multiorbital quantum impurity problems with general interactions and
hybridizations. We show that the method, originally developed to overcome
the dynamical sign problem in certain real-time propagation problems,
can also overcome the sign problem as a function of temperature for
equilibrium quantum impurity models. This is shown in several cases
where the current method of choice, the continuous-time hybridization
expansion, fails due to the sign problem. Our method therefore enables
simulations of impurity problems as they appear in embedding theories
without further approximations, such as the truncation of the hybridization
or interaction structure or a discretization of the impurity bath
with a set of discrete energy levels, and eliminates a crucial bottleneck
in the simulation of ab initio embedding problems.
\end{abstract}
\maketitle

Quantum impurity models describe a small number of strongly interacting
confined states coupled to wide noninteracting baths. While originally
introduced to address magnetic impurities in metals \citep{anderson_localized_1961},
they are now predominantly employed in the context of embedding theories
such as the dynamical mean field theory \citep{metzner_correlated_1989,georges_hubbard_1992,georges_dynamical_1996},
its variants \citep{biermann_first-principles_2003,maier_quantum_2005,held_realistic_2006,kotliar_electronic_2006},
and the self-energy embedding theory \citep{kananenka_systematically_2015,zgid_finite_2017}.
In these theories, the solution of the intractable continuum quantum
many-body system describing a correlated material is approximately
mapped onto a sequence of effective quantum impurity problems coupled
by a self-consistency condition that determines their bath parameters.
Evaluating properties of correlated materials then requires repeatedly
obtaining the Green's functions of quantum impurity problems.

Only in the simplest cases can impurity problems be solved at polynomial
cost. Examples are impurities comprising only a single interacting
orbital, for which powerful continuous-time quantum Monte Carlo (QMC)
\citep{rubtsov_continuous-time_2004,rubtsov_continuous-time_2005,werner_continuous-time_2006,gull_continuous-time_2008,gull_continuous-time_2011}
and renormalization group algorithms \citep{bulla_numerical_2008,wolf_chebyshev_2014,stadler_interleaved_2016}
exist. Other examples are systems with high symmetry and/or special
interactions. For instance, particle\textendash hole symmetry with
local density\textendash density interactions, which allows the solution
of interacting impurity problems with hundreds of sites \citep{leblanc_equation_2013};
or systems where the coupling to the baths (the ``hybridization'')
does not mix the eigenstates of the confined Hamiltonian (\textit{i.e.}
is diagonal in that basis). Everywhere else, the solution of the
impurity model either suffers from a ``sign problem'' that causes
an exponential scaling (as a function of temperature, interaction,
and number of interacting orbitals) or requires additional approximations,
such as the discretization of the continuum of bath states and their
approximation with a set of relatively few discrete bath levels \citep{caffarel_exact_1994,koch_bath_2008,zgid_truncated_2012,lu_efficient_2014,shee_coupled_2019,zhu_coupled_2019}.

In the context of embedding simulations of electronic structure problems,
these limitations are severe. Many important correlated systems contain
transition metal atoms with multiple correlated orbitals. Their symmetries
are rarely high enough that only diagonal hybridization is expected,
especially when surface problems are studied \citep{gorelov_relevance_2009}.
Furthermore, the phenomena of interest often only appear at low temperature.
As a consequence, practitioners typically neglect the terms generating
the sign problem in the Hamiltonian and readjust the remaining parameters
by hand. This limits the predictive power of embedding methods and
their use in ab-initio frameworks, but enables simulations at temperatures
that would otherwise not be accessible. A numerical method able to
reach low temperatures for general impurity Hamiltonians without suffering
from an exponential slowdown would eliminate the need for these approximations
and bridge a central gap in the road to predictive simulations of
correlated electron systems.

In this paper, we present an Inchworm QMC method that overcomes the
low temperature sign problem in multiorbital impurity models, thereby
eliminating these limitations. The method builds on an idea developed
to address real-time dynamics of single orbital quantum impurity problems
\citep{cohen_taming_2015,chen_inchworm_2017,antipov_currents_2017,boag_inclusion-exclusion_2018,dong_quantum_2017,krivenko_dynamics_19},
which overcomes the dynamical sign problem, \emph{i.e.} the exponential
scaling as a function of time, in certain nonequilibrium setups \citep{cohen_taming_2015}.
A key insight in this regard is that both the dynamical and multiorbital
sign problems stem from changing signs in the hybridizations, which
the Inchworm method is able to deal with. We emphasize that the method
does \emph{not} present a general solution to the fermion sign problem
(\textit{i.e.} the exponential scaling as a function of impurity size).

\paragraph{Generic model and method}

We consider generic impurity Hamiltonians of the form $\hat{H}=\hat{H}_{I}+\hat{H}_{B}+\hat{H}_{IB}.$
Here,
\begin{align}
\hat{H}_{I} & =\sum_{ij\sigma}^{N}\varepsilon_{ij,\sigma}\hat{d}_{i\sigma}^{\dagger}\hat{d}_{j\sigma}+\sum_{ijkl\sigma\sigma'}^{N}U_{ijkl}\hat{d}_{i\sigma}^{\dagger}\hat{d}_{k\sigma'}^{\dagger}\hat{d}_{l\sigma'}\hat{d}_{j\sigma}\label{eq:H_impurity}
\end{align}
is a Hamiltonian on a local Hilbert space with $N$ orbitals and local
one-body ($\varepsilon_{ij}$) and two-body ($U_{ijkl}$) terms; 
\begin{align}
\hat{H}_{B} & =\int\mathrm{d}k\varepsilon\left(k\right)\hat{b}^{\dagger}\left(k\right)\hat{b}\left(k\right)\label{eq:H_bath}
\end{align}
is a noninteracting, typically continuous bath Hamiltonian with dispersion
$\varepsilon\left(k\right)$; and
\begin{align}
\hat{H}_{IB} & =\sum_{i}\int\mathrm{d}k\left[t_{i}\left(k\right)\hat{b}^{\dagger}\left(k\right)\hat{d}_{i}+\mathrm{h.c.}\right]\label{eq:H_coupling}
\end{align}
is the impurity\textendash bath coupling Hamiltonian with hopping
terms $t_{i}\left(k\right)$. The $\hat{d}_{i}^{\left(\dagger\right)}$
and $\hat{b}^{\left(\dagger\right)}\left(k\right)$ destroy (create)
particles on the impurity and in the baths, respectively. The computational
challenge consists of computing the single-particle imaginary-time
Green's function $G_{ij}\left(\tau\right)=-\left\langle T_{\tau}\hat{d}_{i}\left(\tau\right)\hat{d}_{j}^{\dagger}\left(0\right)\right\rangle $.

The present method of choice, CT-HYB \citep{werner_continuous-time_2006,werner_hybridization_2006,gull_continuous-time_2011},
proceeds by expanding the partition function $Z=\text{\text{Tr}}e^{-\beta\hat{H}}$
(with $\beta$ the inverse temperature) into a diagrammatic series
in terms of the hybridization, Eq.~\ref{eq:H_coupling}. Using Wick's
theorem, diagrams are combined into determinants \citep{werner_continuous-time_2006}
and stochastically sampled in a random walk procedure \citep{prokofev_polaron_1998,rubtsov_continuous-time_2005}.
Green's functions are measured by eliminating hybridization lines
from partition function diagrams \citep{werner_continuous-time_2006}.
The method scales exponentially in the number of orbitals $N$, as
the local Hamiltonian needs to be diagonalized \citep{werner_hybridization_2006,haule_quantum_2007}.
It scales polynomially in the inverse temperature if the Wick determinants
are positive for each diagram. As this is only the case in certain
high symmetry situations, the algorithm suffers from a sign problem
in general and its scaling is exponential with inverse temperature.

The multiorbital Inchworm method presented in this paper is an imaginary-time
adaptation of the real-time formalism of Ref.~\citep{cohen_taming_2015}
combined with the multiorbital formulation of CT-HYB described in
Ref.~\citep{werner_hybridization_2006}. The fundamental objects
in the method are imaginary-time impurity propagators $Z(\tau)=\text{Tr}_{B}e^{-\tau\hat{H}}$,
where $\text{Tr}_{B}$ denotes a trace over bath states. Propagators
are sequentially obtained on a uniform grid $\tau_{n}=n\Delta\tau$
with integer $n$ increasing from 0 to $\frac{\beta}{\Delta\tau}$.
The first steps (small $n$) are similar to high temperature simulations
and easily performed with CT-HYB. In later steps at larger $n$, the
algorithm efficiently expresses $Z\left(\tau_{n}\right)$ in terms
of $\left\{ \left.Z\left(\tau_{i}\right)\right|0\le i<n\right\} $.

As has been shown in the context of real-time algorithms \citep{cohen_taming_2015},
this incremental procedure vastly reduces the number of diagrams to
be computed and thereby decomposes one large, difficult calculation
into many interdependent, easier ones. The technical implementation
of the method closely follows the real-time implementation described
in detail in Refs.~\citep{cohen_taming_2015}, with two changes.
First, the propagation direction of real time $t$ is replaced by
the orthogonal imaginary time direction $\tau=it.$ Second, the large
value of the propagators and partition functions requires normalization
with $e^{-\tau\hat{H}_{I}}$ in order to avoid numerical instabilities.

After obtaining the propagators, the Green's functions are computed
according to the procedure detailed in Ref.~\citep{antipov_currents_2017}:
$Z(\beta)G_{ij}(\tau)$ is obtained in a separate expansion, and $G_{ij}\left(\tau\right)$
extracted by division with $Z\left(\beta\right).$ Implementation
details and techniques are otherwise identical to Ref.~\citep{antipov_currents_2017},
and fast summation techniques \citep{boag_inclusion-exclusion_2018}
can be used. A brief derivation of the algorithm is included in the
supplemental materials \footnote{See supplemental materials for a brief derivation of the equilibrium
algorithm}.

In order to illustrate the power of the algorithm, we focus on two
setups that are known to be difficult for state-of-the-art algorithms.
Parameters are chosen such that a first set is straightforwardly accessible
with current technology; a second set is difficult but possible; and
a third set is far out of reach of current methods. Due to the interconnected
nature of the Inchworm simulations we choose to compare errors on
observables of interest (such as the Green's function) for a fixed
CPU time in each model. Confidence interval estimates (shaded regions
in all figures) were obtained from a Jackknife analysis of 5 independent
calculations, allowing us to account for nonlinear error propagation
and potential error amplification in the Inchworm algorithm.

\paragraph{Spinless Anderson Model}

\begin{figure}
\includegraphics{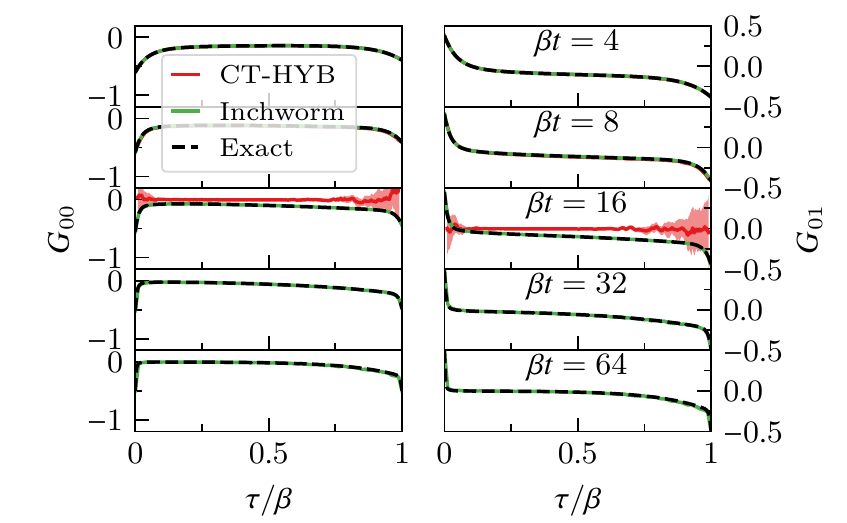}\caption{Imaginary-time Green's function for the spinless Anderson model at
temperatures $T=t/4$ (top panel), $T=t/64$ (bottom panel), and for
intermediate temperatures indicated. Diagonal (left panels) and off-diagonal
(right panels) elements for a small discrete bath (see text) obtained
with exact diagonalization (dashed black), CT-HYB (solid red) and
Inchworm (solid green). CT-HYB data is only available down to $T=t/16$,
and the systematic deviation of the CT-HYB results are indicative
of an additional ergodicity problem.\label{fig:spinless_greens_functions}}
\end{figure}

We first examine the two-orbital spinless Anderson model (SAM) \citep{kashcheyevs_unified_2007,hartle_decoherence_2013},

\begin{equation}
\begin{aligned}\hat{H} & =\sum_{i\in\left\{ 0,1\right\} }\varepsilon_{i}\hat{\text{n}}_{i}+U\hat{\text{n}}_{0}\hat{\text{n}}_{1}-v\left(\hat{d}_{0}^{\dagger}\hat{d}_{1}+\mathrm{h.c}\right)\\
 & +\sum_{k}\sum_{i\in\left\{ 0,1\right\} }\varepsilon_{ik}\hat{b}_{ik}^{\dagger}\hat{b}_{ik}-t\sum_{k}\sum_{i\in\left\{ 0,1\right\} }\left(\hat{d}_{i}^{\dagger}\hat{b}_{ik}+\mathrm{h.c}\right)\\
 & -t^{\prime}\sum_{k}\left(\hat{b}_{0k}^{\dagger}\hat{b}_{1k}+\mathrm{h.c}\right),
\end{aligned}
\label{eq:SAM_Hamiltonian}
\end{equation}
which is a minimal model exposing the exponential scaling issues in
CT-HYB \citep{parcollet_triqs:_2015,seth_triqs/cthyb_2016}. Here
$i$ enumerates the two (spinless) orbitals $0$ and $1$ at local
level energy $\varepsilon_{i}$. Each orbital is connected to its
own bath orbitals (enumerated by quantum numbers $i$ and $k$ and
at level energy $\varepsilon_{ik}$) with coupling $t$ and to the
other orbital with coupling $v$. The density in orbital $i$ is $\hat{n}_{i}=\hat{d}_{i}^{\dagger}\hat{d}_{i}$
and $U$ is a local inter-orbital interaction strength. Pairs of same-$k$
orbitals in the two baths are connected with a hopping $t^{\prime}$.
Here we choose two degenerate, discrete bath states per orbital at
zero energy ($\varepsilon_{ik}=0$ for $i,k=0,1$). The remaining
parameters are set to $\varepsilon_{i}=0$, $U=4t$, $v=t$ and $t^{\prime}=\frac{3}{2}t$.
This finite model can be diagonalized exactly, providing an independent
benchmark for comparison. The mixing of different bath orbitals, here
generated by $t',$ is typical for multiorbital embedding setups and,
in CT-HYB, results in a severe sign problem.

Fig.~\ref{fig:spinless_greens_functions} shows the exact Green's
functions (dashed black line) along with the results from CT-HYB (red)
and Inchworm (green). The parameters were chosen such that the sign
problem in CT-HYB is particularly large \citep{parcollet_triqs:_2015,seth_triqs/cthyb_2016}.
For symmetry reasons $G_{00}=G_{11}$ and $G_{01}=G_{10}$; we therefore
only show $G_{00}$ (left panels) and $G_{01}$ (right panels). Note
that while $G_{00}$ is strictly negative and convex, $G_{01}$ is
neither. Six temperatures ($\beta t=4,8,16,32$, and $\beta t=64$)
are shown, in decreasing order from the top to the bottom panels.
CT-HYB data are only available for $\beta t=4,8,$ and $\beta t=16.$
The time discretization parameter $\Delta\tau$ was set to $\beta/80$,
and the maximum Inchworm diagram order was restricted to 8 \citep{cohen_taming_2015}.
In this simulation, the maximum order was rarely reached and, as also
evidenced by comparison with the exact result, diagram truncation
and discretization errors are smaller than stochastic errors. Inchworm
(CT-HYB) calculations were run for 0.5K (12K) core hours; both algorithms
are trivially parallelizable.

At high temperature (top two panels), all methods agree. Statistical
errors are slightly larger for CT-HYB at the second temperature. However,
at $\beta t=16$, CT-HYB breaks down, exhibiting both large errors
and additional ergodicity issues. For the same parameters, the Inchworm
method remains accurate and consistent with the exact reference, and
results remain correct down to the lowest temperature shown, $\beta t=64$.
This behavior is generic for models with off-diagonal (bath-mixing)
hybridizations.

\paragraph{Kanamori Model}

\begin{figure}
\includegraphics{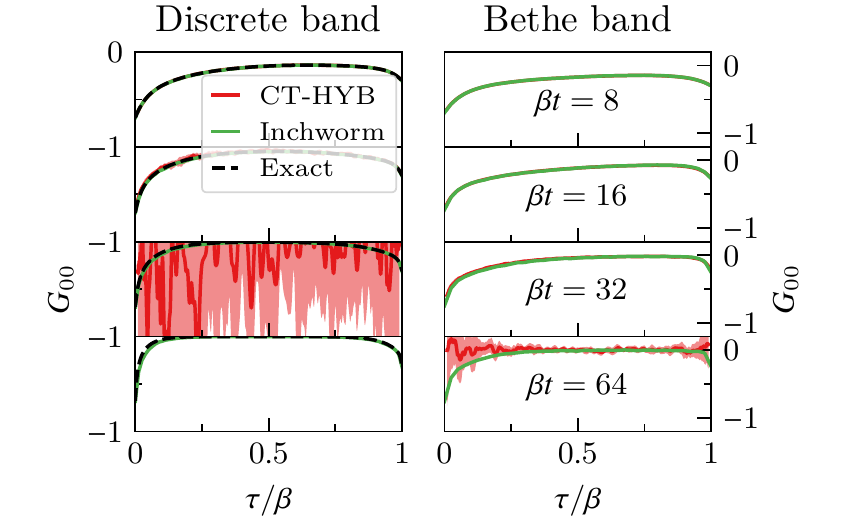}\caption{Diagonal imaginary-time Green's function for the Kanamori model at
temperature $T=t/8$ (top panel), $T=t/64$ (bottom panel), and intermediate
temperatures $T=t/16$ and $T=t/32$, for a discrete band (left panels)
and a semi-circular band with $t=1$. Results from CT-HYB (solid red)
and Inchworm (solid green), along with exact diagonalization where
available (black, left panel only).\label{fig:kanamori_greens_functions}}
\end{figure}

We now consider the two-orbital Kanamori model with spherically symmetric
interactions, which has two orbitals with two spins each. Kanamori
models exhibit interesting non-Fermi-liquid ``spin freezing'' \citep{werner_spin_2008}
and ``Hund's metal'' physics \citep{yin_kinetic_2011,medici_janus_2011,georges_strong_2013},
and are frequently considered in multiorbital DMFT simulations of
transition metal compounds with cubic symmetry. The two-orbital variant
with local Hamiltonian
\begin{equation}
\begin{aligned}\hat{H} & =U\sum_{i\in\left\{ 0,1\right\} }\hat{n}_{i\downarrow}\hat{n}_{i\uparrow}+\left(U-2J\right)\sum_{i\neq j}\hat{n}_{i\downarrow}\hat{n}_{j\uparrow}\\
 & +\left(U-3J\right)\sum_{i>j,\sigma}\hat{n}_{i\sigma}\hat{n}_{j\sigma}\\
 & +J\sum_{i\neq j}\left(\hat{d}_{i\uparrow}^{\dagger}\hat{d}_{j\downarrow}^{\dagger}\hat{d}_{i\downarrow}\hat{d}_{j\uparrow}+\mathrm{h.c.}\right)
\end{aligned}
\label{eq:kanamori_hamiltonian}
\end{equation}
is commonly used for $e_{g}$ bands in correlated 3d or 4d orbitals.
The local part of the model is parameterized by the Coulomb $U$ and
Hund's $J$ interaction parameters, which we set to $U=2t$ and $J=0.2t$.
The ``spin-exchange'' and ``pair-hopping'' terms in the last line
are the only non-density\textendash density terms and\textemdash although
frequently neglected\textemdash have an important effect on the physics
\citep{werner_spin_2008}. Within DMFT, this local Hamiltonian hybridizes
with a bath modeled by a frequency- and orbital-dependent hybridization
function $\Delta_{ij}(\omega)$ that generically mixes different orbitals.

In the following, we consider two hybridization functions $\Delta_{ij}=\left(\delta_{ij}+r\left(1-\delta_{ij}\right)\right)t^{2}\mathcal{G}\left(\omega\right)$,
where $r$ controls the relative size of off-diagonal elements and
$a$ controls the overall coupling strength. The first hybridization
is an exactly solvable discrete band $\mathcal{G}\left(\omega\right)=\sum_{k}\delta\left(\omega-\varepsilon_{k}\right)$
with two levels $\varepsilon_{k}\in\left\{ \pm2.3t\right\} $ per
spin-orbital, at $r=\frac{1}{2}$ and $a=1$. The second hybridization
describes coupling to a continuous semicircular band $\mathcal{G}\left(\omega\right)=\frac{2}{\pi D^{2}}\sqrt{D^{2}-\omega^{2}}$,
where the half bandwidth is $D$. Here we set $r=1$, $D=2t$ to consider
a band as it occurs in the solution of the dynamical mean field equations
in the infinite coordination number limit on a Bethe lattice.

Fig.~\ref{fig:kanamori_greens_functions} shows the diagonal, same-spin
Green's function elements, $G_{00}\left(\tau\right)\equiv G_{i\sigma,i\sigma}\left(\tau\right)$.
The discrete case (left panels) can be exactly diagonalized (dashed
black curves), but for the continuous case (right panels) no analytical
results are available. The sign problem in this system is not as severe
as in the SAM, and we can therefore present CT-HYB results down to
half the lowest temperature in Fig.~\ref{fig:spinless_greens_functions}.
The numerical parameters and statistical analysis are as in the SAM.
All Inchworm (CT-HYB) calculations were run for 1.5K (3K) core hours.

While CT-HYB performs reasonably well at high temperature, it breaks
down for both band types as $T$ is lowered to $\beta t=32$ (left
panels) and $\beta t=64$ (right panels). Inchworm shows controlled
results for all cases in both models, though small deviations between
Inchworm and the exact solution, due to discretization errors, are
visible in the bottom left panel. We verified that these deviations
can easily be removed by decreasing $\Delta\tau$ (not shown).

\paragraph{Scaling analysis}

\begin{figure}
\includegraphics{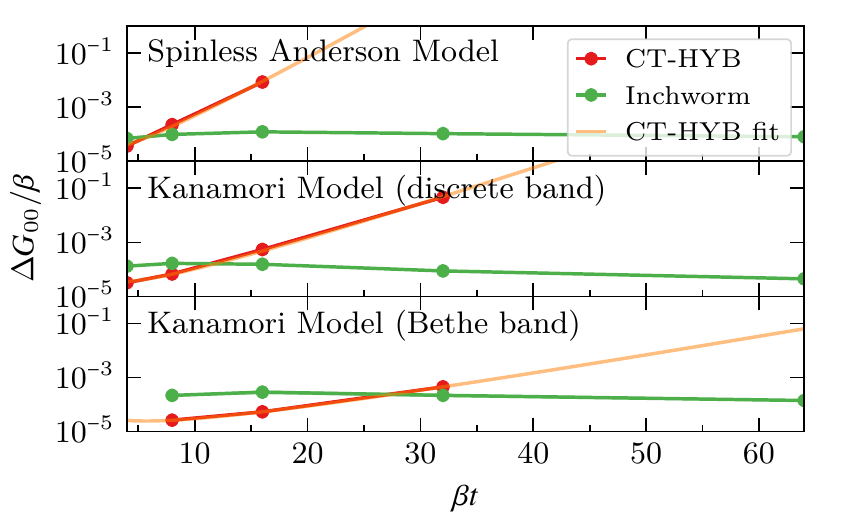}\caption{Scaling analysis. Mean absolute deviation from exact result (top two
panels, SAM and discrete Kanamori models) and statistical standard
errors (bottom panel, continuous Bethe band Kanamori model) divided
by inverse temperature $\beta$, as a function of $\beta$. CT-HYB
(red) and Inchworm (green) results are shown along with exponential
fits for CT-HYB (light red).\label{fig:scaling}}
\end{figure}

The results presented so far are qualitative, insofar as CT-HYB breaks
down in several cases where Inchworm does not. To gain additional
insight into the nature of the observed breakdown and the relative
merits of the two methods, we present a quantitative error analysis
in Fig.~\ref{fig:scaling}. We plot Green's function error estimates
for the three different models as a function of the inverse temperature
$\beta$. The errors are divided by $\beta$ and shown on a logarithmic
scale. For the SAM and the discrete Kanamori models (two upper panels),
the error estimates are given by the absolute value of difference
from the exact result, $\Delta G\equiv\left|G-G_{\mathrm{\text{Exact}}}\right|$;
they therefore also take into account any systematic bias the Monte
Carlo methods might exhibit. In the lower panel, for the Kanamori
model coupled to a continuous Bethe band, no exact result is available.
The errors are therefore obtained from a Jackknife analysis on 5 independent
calculations, and account only for the magnitude of variation between
these runs. We note that the absolute value of the errors is of course
implementation dependent. Here we used the highly optimized ALPS CT-HYB
code \citep{shinaoka_continuous_2017,gaenko_updated_2017,wallerberger_alpscore_2018},
and an Inchworm implementation written in C++.

In all cases shown, ALPS CT-HYB is more accurate than our Inchworm
implementation for high temperature. However, as a function of $\beta,$
the $\frac{\Delta G}{\beta}$ obtained within CT-HYB is at least exponential
in $\beta$ (see fits in Fig.~\ref{fig:scaling}, which are extrapolated
beyond where CT-HYB errors can be reliably obtained). This exponential
scaling is a consequence of the presence of a sign problem. In contrast,
the $\frac{\Delta G}{\beta}$ obtained with the Inchworm method is
essentially flat, implying a \emph{linear scaling} in inverse temperature.
This means that, for the systems presented, the Inchworm method presents
a solution to the sign problem as a function of temperature and allows
access to temperatures that are much lower than what is possible with
CT-HYB. For example, Fig.~\ref{fig:scaling} shows that in the Bethe
case, obtaining a result of comparable quality to our Inchworm data
at $\beta=64$ with ALPS CT-HYB would take \textasciitilde$3\times10^{9}$
core hours or \textasciitilde 342K core years.

The linear scaling of the Inchworm method should be interpreted only
as a lower bound: at even lower temperatures, a finer time discretization
or a generalization to a non-uniform grid will be needed to maintain
accuracy. We expect this to result in a low-order (but more than linear)
polynomial scaling in the inverse temperature \footnote{See supplemental materials for a discussion of the scaling}.
We emphasize again that the method is not a general solution of the
fermion sign problem, as it remains explicitly exponential in the
number of interacting orbitals.

\paragraph{In conclusion,}

we present an equilibrium multiorbital quantum impurity solver based
on the Inchworm method. We show, for two generic scenarios, that the
method avoids the exponential scaling with inverse temperature observed
in other methods and thereby presents a ``solution'' to this particular
class of sign problems. A comparison to the state-of-the-art method,
CT-HYB, shows that parameter regimes are now accessible that were
previously out of reach of numerically exact quantum impurity solvers.

Our Inchworm impurity solver addresses a critical need for numerically
exact multiorbital quantum impurity solvers that can treat both generic
four-fermion interaction terms and generalized non-diagonal hybridization
functions. This need stems from embedding constructions such as the
DMFT or the self-energy embedding theory, where hybridization functions
typically arise as continuous multiorbital functions in frequency
space and interactions are not of the density\textendash density type.
By being able to solve such impurity problems without introducing
additional artificial discretizations of the bath orbitals, and without
further truncating and approximating the hybridization and interaction
structure, the method bridges an important gap on the route to controlled
ab-initio many-body embedding theories.
\begin{acknowledgments}
G.C. acknowledges support by the Israel Science Foundation (Grant
No.~1604/16). E.E. and G.C. acknowledge support by the PAZY foundation
(Grant No.~308/19), and E.G. was supported by DOE ER 46932. Computational
support was provided by the NegevHPC project \citep{noauthor_negevhpc_nodate}.
International exchange and collaboration was supported by Grant No.~2016087
from the United States-Israel Binational Science Foundation (BSF).
\end{acknowledgments}

\bibliographystyle{apsrev4-1}
\bibliography{Library}

\end{document}